\documentclass[prb,twocolumn,floatfix,showpacs,10pt,longbibliography]{revtex4-2}
\usepackage{graphicx}
\usepackage{dcolumn}
\usepackage{bm}
\usepackage{subfigure}
\usepackage{blindtext}
\usepackage{amsmath}
\usepackage{units}
\usepackage{float}

\usepackage[mathlines]{lineno}
\usepackage{xcolor}

\begin{document}
\title{Improved imaging of magnetic domains with a photoelectron emission microscope by utilizing symmetry and momentum selection}

\author{F. O. Schumann}
\altaffiliation{Electronic mail: schumann@mpi-halle.de}
\affiliation{ Max-Planck-Institut f\"{u}r Mikrostrukturphysik, Weinberg 2, 06120 Halle, Germany}

\author{M. Paleschke}
\author{J. Henk}
\author{W. Widdra}
\affiliation{Institute of Physics, Martin-Luther-Universit\"{a}t Halle-Wittenberg, Von-Danckelmann-Platz 3, D-06099 Halle (Saale), Germany}

\author{C.-T. Chiang}
\affiliation{Institute of Atomic and Molecular Sciences, Academia Sinica, Taipei, Taiwan}

\date{\today}

\begin{abstract}
Imaging of magnetic domains with a photoelectron emission microscope operated with photon energies in the threshold regime often suffers from low contrast. In this work we show by symmetry considerations, photoemission calculations, and imaging experiments, how the contrast can be improved significantly. The key to both domain selectivity and sizable intensity asymmetries is, guided by symmetry considerations, selecting the momenta of the photoelectrons by a properly positioned contrast aperture. By comparing computational with experimental results for an Fe(001) surface we prove the feasibility of the approach. 
\end{abstract}

\pacs{73.20.At, 79.60.-i}

\maketitle

 
\section{Introduction}
A striking consequence of the exchange interaction, which itself originates from the Coulomb interaction between electrons and the Pauli principle, is ferromagnetism. It causes parallel alignment of the electron spins, leading to the macroscopic observable magnetization. The magnetization is oriented along main symmetry directions of the crystal. This magnetic anisotropy is brought about by the spin-orbit interaction. Hence, a complete theoretical description of the  electronic valence states of ferromagnets must include both the exchange and the spin-orbit interaction. Experimentally, access to the electronic structure is possible via spin- and angle-resolved photoemission.

The presence of both spin-orbit coupling and exchange interaction leads to the phenomenon of magnetic dichroism in the absorption of circularly polarized electromagnetic radiation at core levels. This X-ray magnetic circular dichroism (XMCD) is element-specific and  allows to determine the orbital and spin magnetic moments \cite{347_Schuetz,85_Chen}. The effect of magnetic dichroism also shows up in the spin-integrated photoemission spectra from core levels and valence states \cite{40_Baumgarten,145_Getzlaff,2359_Kuch,2422_Kuch,2368_Bansmann,2388_Venus,2293_Kuch}. Via threshold photoemission it is also possible to achieve magnetic contrast. 

The appeal of threshold dichroism is the possibility to control the polarization state of the light at these low photon energies, without the need to employ synchrotron radiation. Utilizing a photoelectron emission microscope (PEEM), it is possible to obtain domain images \cite{2290_Feder,2374_Marx,2414_Nakagawa,2285_Kronseder,2464_Hild, 2437_Melchior}.

\begin{figure}
	\includegraphics[width = \columnwidth]{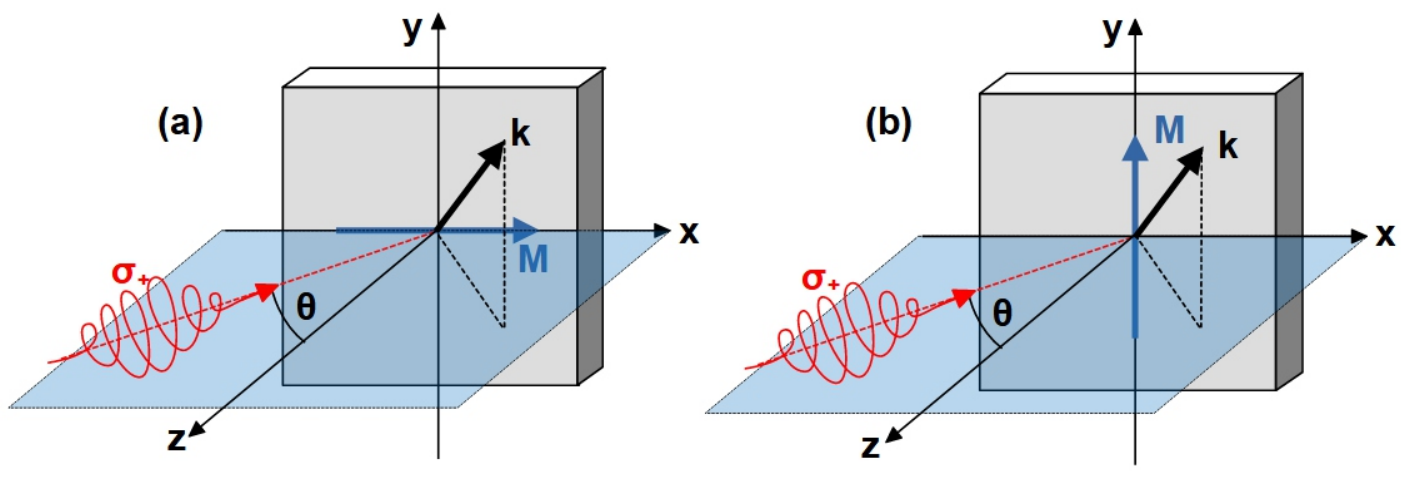}
	\caption{Sketch of the experimental geometry. The incoming circularly polarized light defines the scattering plane, indicated in light blue. The angle of incidence $\theta$ is defined with respect to the surface normal ($z$ axis).  The emitted electron is characterized by the momentum vector $\vec{k}$. The magnetization directions are in-plane and either along the $x$- (a) or the $y$ axis (b). \label{geometry_setup}}
\end{figure}

\begin{table*}
	\caption{\label{tab:table2}
		Effect of $C_{2v}$ symmetry operations on the in-plane components of the momentum, helicity of the light, and the magnetization components \cite{2369_Henk}. The identity operation is $E$, while $\sigma_{xz}$ and $\sigma_{yz}$ stand for the reflections at the $xz$ and the $yz$ plane, respectively. $C_{2}$ is the $\pi$-rotation about the $z$ axis.}
	\begin{ruledtabular}
		\begin{tabular}{cccccccc}
			& \multicolumn{2}{c}{wave vector} & \multicolumn{2}{c}{helicity} & \multicolumn{3}{c}{magnetization} \\
			\hline
			E&$k_x$ & $k_y$ &$\sigma_{+}$ &$\sigma_{-}$ & $M_x$ & $M_y$ &$M_z$ \\
			$\sigma_{xz}$& $k_x$ & -$k_y$ &$\sigma_{-}$ &$\sigma_{+}$ &-$M_x$ & $M_y$ & -$M_z$ \\
			$\sigma_{yz}$&-$k_x$ & $k_y$ &$\sigma_{-}$ &$\sigma_{+}$ &$M_x$ & -$M_y$ &-$M_z$ \\
			$C_2$&-$k_x$&- $k_y$ &$\sigma_{+}$ &$\sigma_{-}$ & -$M_x$ & -$M_y$ & $M_z$ \\
			
		\end{tabular}
	\end{ruledtabular}\label{c2v_table}
\end{table*}

At the heart of the magnetic circular dichroism effect are four fundamental intensities, one for each combination of magnetization orientation and circular polarization state of the light. These allow to define three asymmetries and the total intensity \cite{2369_Henk}. For domain imaging, however, it turns out  advantageous to define a new set of parameters. We identify the terms relevant for domain contrast below.

The crystal symmetry imposes relations among the fundamental intensities which carry over to the asymmetries. With an appropriate selection of the emission directions one can utilize the symmetry for selective domain contrast for in-plane magnetization. In order to visualize these points, we have performed  numerical calculations of the photoemission process from an Fe(001) surface via a relativistic one-step description \cite{2430_Halilov,2419_Feder,2412_Henk,2343_Scheunemann,2290_Feder,2369_Henk}. We focused at low photon energies in the threshold regime, which allows domain imaging by a photoelectron emission microscope in combination with a discharge lamp or ultraviolet (UV) laser radiation and accompanying additional polarization optics. The proposed domain selectivity was experimentally confirmed by measurements on an Fe(001) surface.

This paper is organized as follows. In section~\ref{sec:asymmetries} we identify asymmetries suitable for domain imaging. Details on experiments and theory are presented in section~\ref{sec:details}, while symmetry is considered in section~\ref{sec:symmetry}. In section~\ref{sec:numerical-asymmetries} we apply the asymmetries to computed spectra for Fe(001); their improvement by momentum selection is shown in section~\ref{sec:improving}. That our approach can be applied to perpendicular magnetized samples is proven in section~\ref{sec:pma}, followed by our  conclusion in section~\ref{sec:conclusion}.

\section{Asymmetries for domain imaging}
\label{sec:asymmetries}
Conceptually, a complete spin-integrated measurement requires the determination of four intensities $I_{\sigma, M}$, in which the polarization state of the light $\sigma$ is switched between positive ($\sigma_+$) and negative helicity ($\sigma_-$). Likewise the magnetization orientation is switched between $M_{+}$ and $M_{-}$. These four intensities can be arithmetically regrouped into three differences and the sum~$S$ (see Appendix~\ref{sec:relations}). The ratio of the differences and the sum then leads to three asymmetries $A_{\mathrm{pol}}$, $A_{\mathrm{mag}}$, and $A_{\mathrm{ex}}$  \cite{2369_Henk}. The subscript  ``pol'' refers to the magnetization averaged (non-magnetic) dichroism, while ``ex''  identifies the exchange part of the dichroism. The dichroic signal obtained with unpolarized light can be identified with ``mag''  \cite{2369_Henk}. 

The above asymmetries are appropriate for spectroscopic analyses. In domain imaging, however, the magnetization directions are fixed and only the helicity can be varied. Therefore, each domain produces two intensities rather than four. Hence, it is beneficial to determine the intensity difference and intensity sum of two oppositely magnetized domains. Assuming that $A_{\mathrm{mag}}$ is small, which we will see to be true in the present case, the asymmetries of the two oppositely magnetized domains become
\begin{subequations}
\begin{align} 
	A_{+} & = (A_{\mathrm{pol}}+A_{\mathrm{ex}}) (1-A_{\mathrm{mag}}),
	\label{A_pos}
    \\
     A_{-} & = (A_{\mathrm{pol}}-A_{\mathrm{ex}})(1+A_{\mathrm{mag}})
	\label{A_neg}
\end{align}    
\end{subequations}
(we derive the relations between the two sets of asymmetries in Appendix~\ref{sec:relations}). This means that the asymmetry associated with a  ``+'' or  ``-'' domain is mainly given by the asymmetries $A_{\mathrm{pol}}$ and $A_{\mathrm{ex}}$, while $A_{\mathrm{mag}}$ enters as a minor correction. Further, if $A_{+}$ and $A_{-}$ have the same sign,  $A_{\mathrm{pol}}$ is larger in magnitude than $A_{\mathrm{ex}}$. As we will see below, this is the case for the numerical evaluation of the Fe(001) surface in the threshold regime.

Concerning domain imaging it is important that these two asymmetries $A_{+/-}$ differ significantly. Hence, it is useful to inspect the difference \begin{align} 	 
	A_{\mathrm{diff}} & = A_{+} - A_{-} \approx 2 \, A_{\mathrm{ex}}.
	\label{A_diff}
\end{align}
The key point is that for the domain contrast mainly twice the exchange asymmetry $A_{\mathrm{ex}}$ is relevant. 

\section{Theoretical and experimental details}
\label{sec:details}
An integral part of a photoelectron emission microscope (PEEM) is the contrast aperture at the back focal plane, at which a $\vec{k}_{\parallel}$ image of the intensity is created. The size of  the $\vec{k}_{\parallel}$ image scales with the kinetic energy as $\sqrt{E_{\mathrm{kin}}}$. Therefore, the contrast aperture confines emission angles $\delta$ via the relation $k \cdot \sin(\delta)$. The main purpose of the aperture is to reduce the contributions of the aberrations of the real space image on the detector \cite{2086_Bauer}. A PEEM instrument can be operated in two modi: either a real-space image is mapped on the channel-plate detector or a momentum image is visualized \cite{1089_Kotsugi,1319_Kroemker}.
  
Experimental data have been obtained with a commercial PEEM instrument \cite{2337_Merkel,2613_Paleschke} and a light source with polarization optics and a photon energy of  $h \nu = \unit[5.20]{eV}$. In its standard version it is equipped with five circular contrast apertures, that can be moved within the back focal plane. Under normal operating conditions their sizes converted into momentum radii read $0.051, 0.12, 0.26, 0.86$, and $\unit[3.0]{\AA{^{-1}}}$.

The angle of incidence is $\theta = \unit[65]{^\circ}$, a mirror within the spectrometer allows an additional near normal incidence of $\theta = \unit[4]{^\circ}$, which is approximated as  $\unit[0]{^\circ}$ in the calculations. Energy discrimination is possible via a retarding grid in front of the channel-plate detector. The energy resolution is $\unit[100]{meV}$. All measurements were performed at room temperature.

For the photoemission calculations we utilize a relativistic one-step framework which treats spin-orbit and exchange interaction on equal footing \cite{2430_Halilov,2419_Feder,2412_Henk,2343_Scheunemann,2290_Feder,2369_Henk}. The electronic-structure calculations gave a work function of $\unit[4.63]{eV}$ in good agreement with experimental results for an Fe(001) surface \cite{1195_Bertacco,2478_Cameron,2105_Derry,2477_Kawano}.

\section{Symmetry considerations}
\label{sec:symmetry}
In the following we discuss the symmetry for in-plane easy axes of the magnetization oriented along a high-symmetry direction, as is the case for ferromagnetic Fe(001) and Fe(110) surfaces. For this purpose we adopt a setup as sketched in Fig.~\ref{geometry_setup}. The incoming light and the surface normal define the $xz$-plane as scattering plane, which is chosen to be one of the symmetry planes of a $C_{2v}$-symmetric surface. The polar angle of incidence is given by $\theta$. The emitted electrons are characterized by their momentum $\vec{k}$, which is not confined to the scattering plane, but can have any direction. The magnetization is in-plane: if oriented along $+x$ ($-x$) it is labeled ``right'' (``left''). If the magnetization is perpendicular to the scattering plane, we adopt the notations  ``up'' and  ``down''. 

The symmetry operations of the point group $C_{2v}$ transform the in-plane components of the momentum, helicity of the light and magnetization components as listed in Table~\ref{c2v_table}~\cite{2369_Henk}. For off-normal incidence of the light only the mirror operation at the $\sigma_{xz}$ plane (scattering plane) needs to be considered. 

The in-plane momentum of the emitted electron transforms as $(k_x, k_y) \to (k_x, -k_y)$ , while the helicity is reversed ($\sigma_{\pm} \to \sigma_{\mp}$).  As far as the magnetization is concerned, $M_x$ changes sign, while $M_y$ is unaffected. In Figure~\ref{geometry_setup}  the in-plane magnetization directions are either along the $x$- or the $y$ axis. These axes coincide with the in-plane easy axes of the magnetization for Fe(001) which in the bulk or for films thicker than $\unit[5]{ML}$ lies within the surface plane \cite{2011_Ballentine, 312_Qiu,541_Berger}.

Combining the above results leads to the asymmetry relations
\begin{subequations}
\begin{align} 
       A_{\mathrm{right/left}}(k_x,k_y) & = - A_{\mathrm{left/right}}(k_x,-k_y), \\
	        A_{\mathrm{up/down}}(k_x,k_y) & = -A_{\mathrm{up/down}}(k_x,-k_y)
	\label{asym_trans}
\end{align}    
\end{subequations}
for samples either magnetized ``left/right'' or ``up/down''.  The key point is that for the magnetization being along the $x$ axis (``left''/``right'') the symmetry operation connects opposite magnetization directions, while for the $y$ axis (``up''/``down'') it does not. A consequence for the $y$-aligned magnetization is a sign reversal of the asymmetry upon changing $k_y \to -k_y$. 

The mirror operation yields the relations
\begin{subequations}
\begin{align} 
		A_{\mathrm{pol}}(k_x,k_y) & =-A_{\mathrm{pol}}(k_x,-k_y),  \label{asym_trans2a}\\
		A_{\mathrm{mag}}(k_x,k_y) & =\mp A_{\mathrm{mag}}(k_x,-k_y), \label{asym_trans2b}\\
		A_{\mathrm{ex}}(k_x,k_y)& = \pm A_{\mathrm{ex}}(k_x,-k_y)  \label{asym_trans2}
\end{align}    
\end{subequations}
for the spectroscopic asymmetries. While the behavior of $A_{\mathrm{pol}}$ is the same for $x$- and $y$-aligned magnetization, one has to discriminate for $A_{\mathrm{mag}}$ and $A_{\mathrm{ex}}$.  In Equations~\eqref{asym_trans2b} and \eqref{asym_trans2}, the $+$ sign holds for $x$-magnetization, while for the $y$-magnetization the $-$ sign is valid.

\section{Computed asymmetries of Fe(001)}
\label{sec:numerical-asymmetries}
We now apply the above ideas to domain imaging of Fe(001), adopting the experimental geometry ($\theta = \unit[65]{^{\circ}}$, $h \nu = \unit[5.20]{eV}$; cf.\ Fig.~\ref{geometry_setup}) and focusing on  emission from the Fermi energy $E_{\mathrm{F}}$.

\begin{figure*}
	\includegraphics[width  = \textwidth]{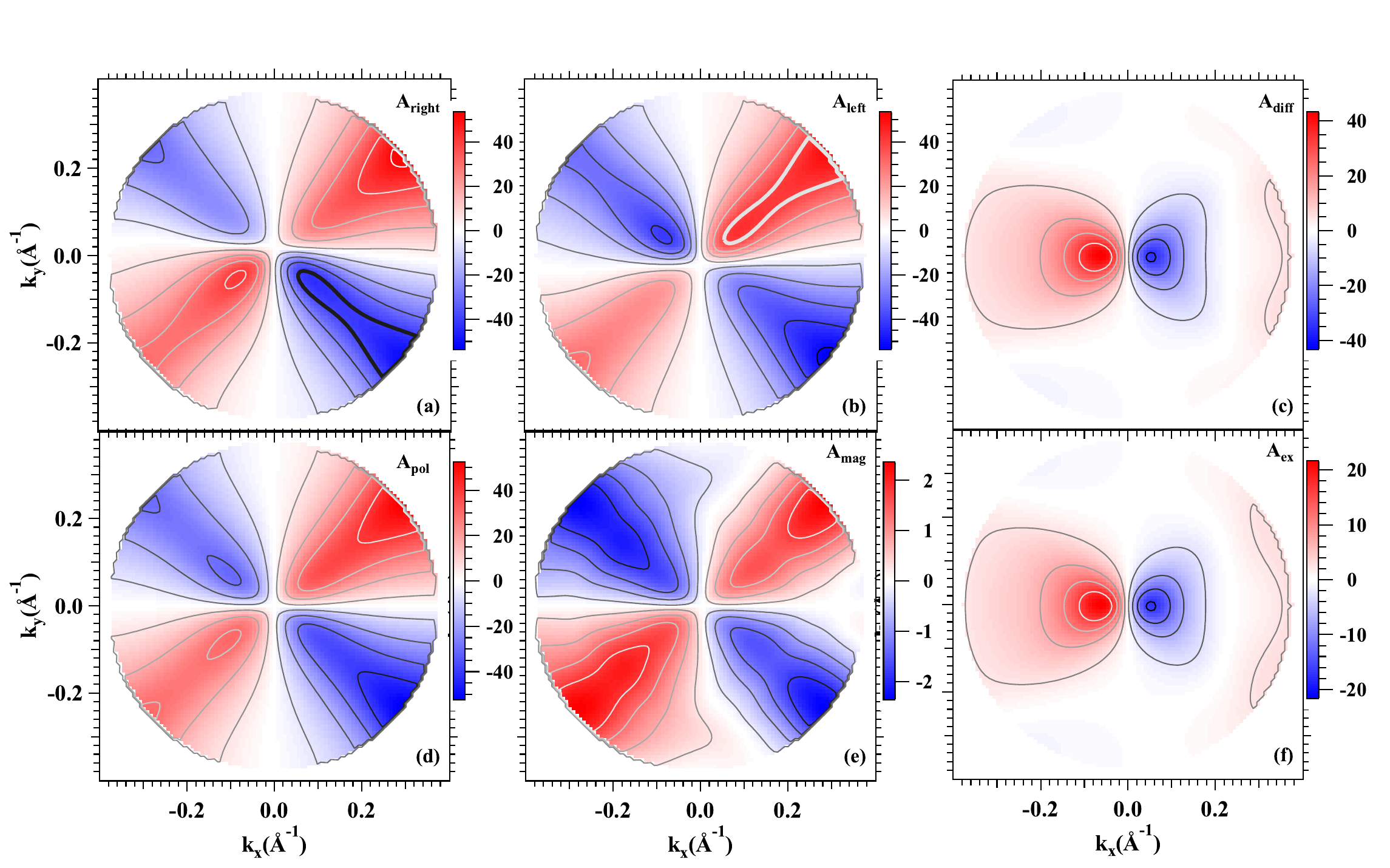}
	\caption{Calculated asymmetries for an Fe(001) surface with magnetization along $\pm x$. The photon energy is set to $h \nu = \unit[5.20]{eV}$ and  $\theta = \unit[65]{^\circ}$. Electron emission from $E_{\mathrm{F}}$ is considered. Contour lines help to identify the action of the symmetry operations. Top row: asymmetries relevant for domain imaging. Bottom row: asymmetries relevant for spectroscopy as explained in the Appendix. The color bar refers to asymmetries in percent. \label{allx}}
\end{figure*}

In Figure~\ref{allx} we present computed results for the magnetization direction being parallel or antiparallel to the $x$ axis (``left''/``right''). The asymmetries $A_{\mathrm{right}}$ and $A_{\mathrm{left}}$, shown in panels (a) and (b), change sign upon moving from one quadrant to the next. The asymmetry values reach absolute values up to $\unit[45]{\%}$. The largest absolute values are taken roughly in the center of each quadrant, while near the $k_x$ and $k_y$ axes the contributions have a nodal line. Further, we note that $A_{\mathrm{right}}$ and $A_{\mathrm{left}}$ are rather similar in shape: in each quadrant the asymmetry's sign is the same and the absolute values are comparable. 

Apparently, the distributions of $A_{\mathrm{left}}$ and $A_{\mathrm{right}}$ closely resemble $A_{\mathrm{pol}}$ in Fig.~\ref{allx}(d), which, recalling Eqs.~\eqref{A_pos} and \eqref{A_neg}, is not surprising. Obviously the contribution of $A_{\mathrm{ex}}$ must be smaller than $A_{\mathrm{pol}}$. More precisely, at regions of large $A_{\mathrm{pol}}$ values, those of $A_{\mathrm{ex}}$ are less. Sizeable $A_{\mathrm{ex}}$ (up to $\unit[20]{\%}$) appears near the $k_x$ axis.

Aiming at domain imaging, we need to identify where the difference between these two distributions is largest. For this purpose we compute $A_{\mathrm{diff}} = A_{\mathrm{right}} - A_{\mathrm{left}}$ (Fig.~\ref{allx}(c)), which exhibits values of up to $\unit[40]{\%}$ near the $k_x$ axis. Finally, the patterns of $A_{\mathrm{diff}}$ and $A_{\mathrm{ex}}$ are similar, apart from a factor of two in the asymmetry values. This finding is in line with the approximation applied in Eq.~\eqref{A_diff}, which is valid for small $A_{\mathrm{mag}}$. We point out that the $A_{\mathrm{diff}}$ pattern in Fig.~\ref{allx}(c) has been determined from the difference of the patterns reproduced in (a) and (b), rather than using the approximation of Eq.~\eqref{A_diff}. Moreover, $A_{\mathrm{ex}}$ and $A_{\mathrm{diff}}$ are similarly distributed, as expected from Eq.~\eqref{A_diff}. This equation was obtained under the assumption that $A_{\mathrm{mag}}$ is small, which is supported by Fig.~\ref{allx}(e).  

We now address the symmetries of the asymmetry distributions in the reciprocal space. In  Figures~\ref{allx}(a) and (b) thick contour lines define an area: black contour in the fourth quadrant in Fig.~\ref{allx}(a), white contour in Fig.~\ref{allx}(b) in the first quadrant. The shape is the same in the two diagrams, but the $k_y$ position has been mirrored at the $k_x$ axis when going from panel~(a) to (b). As a consequence of Eq.~\eqref{asym_trans} the sign has changed as well, while maintaining the absolute value.  For $A_{\mathrm{pol}}$ and $A_{\mathrm{mag}}$ we notice a sign reversal upon mirroring at the $k_x$ axis, while for $A_{\mathrm{ex}}$ and $A_{\mathrm{diff}}$ the value stays the same, which is in line with Eq.~(\ref{asym_trans2}).

\begin{figure*}
	\includegraphics[width = \textwidth]{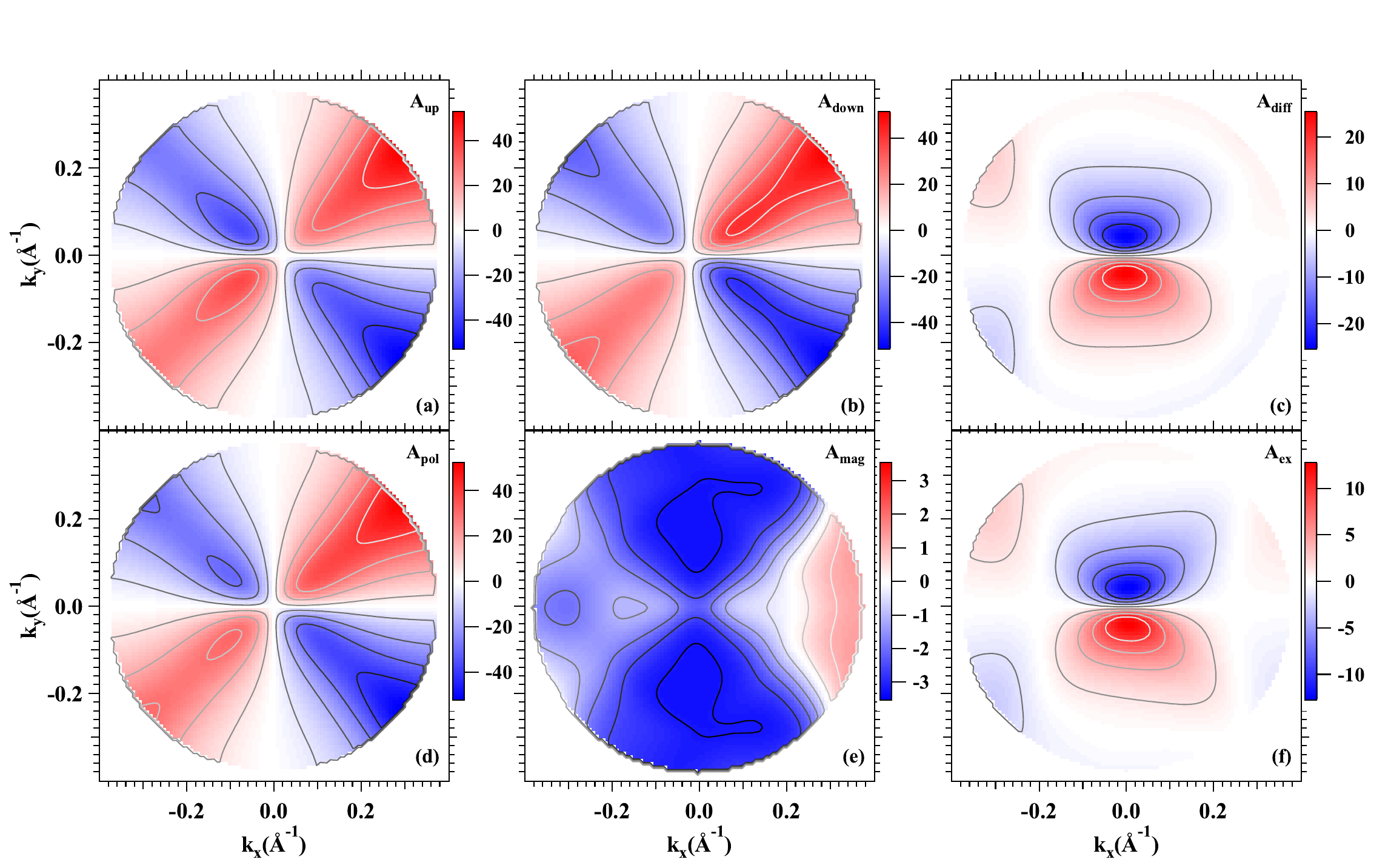}
	\caption{Calculated asymmetries for an Fe(001) surface. The photon energy is set to $h\nu = \unit[5.20]{eV}$ and  $\theta = \unit[65]{^\circ}$.  Electron emission  from $E_{\mathrm{F}}$ is considered. The asymmetries are given in percent. The distribution for $A_{\mathrm{mag}}$ in (e) keeps the sign and absolute value upon reflection at the $x$ axis. For all others the absolute value of the asymmetry stays the same, but changes sign. \label{ally}}
\end{figure*}

Next we discuss results for $y$-aligned magnetization (Fig.~\ref{ally}). As in the case for $x$-magnetization we note alternating signs when moving the contrast aperture from one quadrant to the next (panels~(a) and (b)). Again, the asymmetry values are of the same order of magnitude and very similar to $A_{\mathrm{pol}}$ shown in Fig.~\ref{ally}(d). A consequence of the rule formulated in Eq.~\eqref{asym_trans} is that the asymmetry for the two opposite $y$-magnetization is antisymmetric with respect to reflection at the $k_x$ axis, in contrast to $x$-magnetization.

The different behavior upon reflection at the $k_x$ symmetry line  has an important implication for domain imaging. As mentioned earlier, a contrast aperture of the actual instrument defines a circular region in momentum space which could be positioned arbitrarily. For example the center of this circular region could be located on the $k_x$ axis. In this case the asymmetry values for the $y$-magnetization vanish, hence $A_{\mathrm{diff}} = 0$ and there is no domain contrast. This is in contrast to $x$-magnetization, for which $A_{\mathrm{diff}}$ is non-zero. Put another way, a suitable position of the contrast aperture allows for domain selectivity. We now explore this feature in more detail.

\section{Improving domain imaging with a contrast aperture}
\label{sec:improving}
\begin{figure*}
	\includegraphics[width = \textwidth]{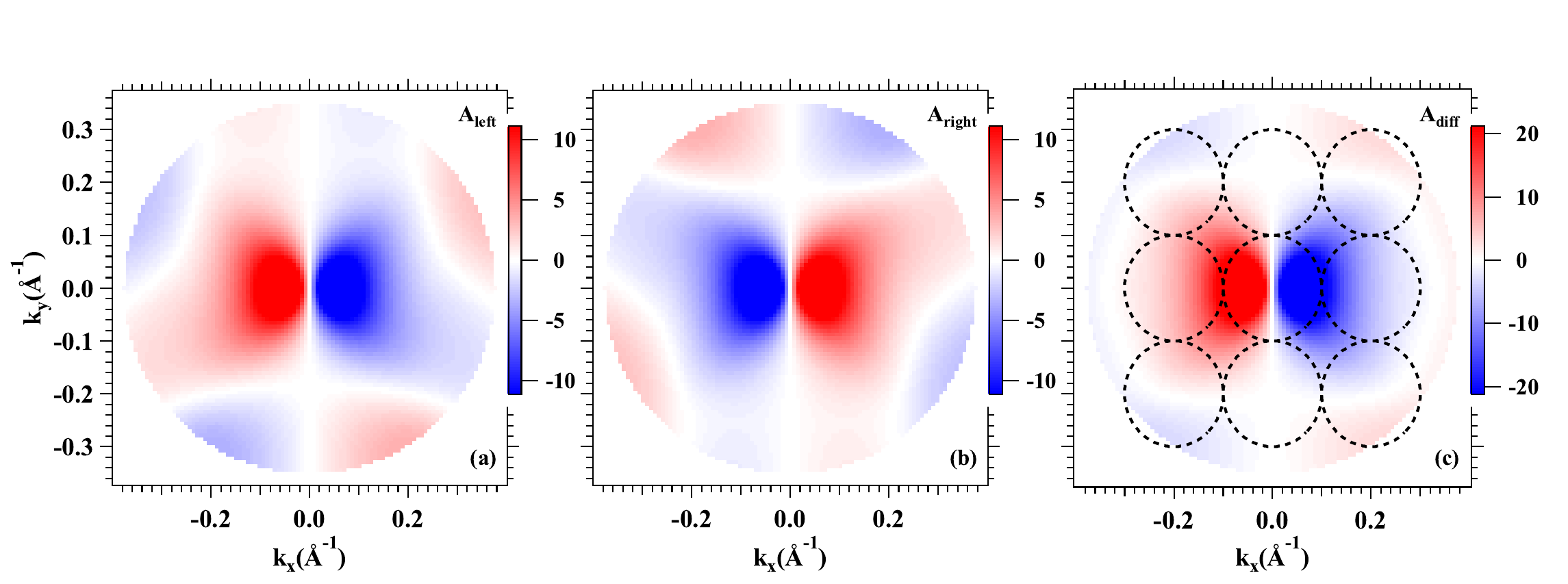}
	\caption{Asymmetry distributions of the two orientations of the x-magnetization (``left''/``right'') in (a) and (b). In (c) we show the difference $A_{\mathrm{diff}}$. Data are calculated for normal incidence and $h \nu = \unit[5.20]{eV}$. Electrons are emitted from $E_{\mathrm{F}}$. Circles in (c) mark momentum ranges with a diameter of $\unit[0.2]{\AA^{-1}}$, given by the position of the contrast aperture.  \label{k_selection}}
\end{figure*}

In the case of normal incidence, an additional symmetry element exists, namely the reflection at the $yz$-plane, whose consequence can be read from Fig.~\ref{k_selection}. There we show the asymmetries for $x$-magnetization (``left''/``right'') and their difference $A_{\mathrm{diff}}$ (we kept the photon energy of $\unit[5.20]{eV}$ and select electrons emitted from $E_{\mathrm{F}}$). The additional $yz$-symmetry plane shows up as a zero-crossing  upon reflection at the $k_y$ axis and as sign reversal of the asymmetries while maintaining the amplitude. This behavior is identical to that of the asymmetries shown in Fig.~\ref{ally}(a)-(c) for $y$-magnetization.

Considering the symmetry arguments, it is evident that for an aperture centered at $k_x = k_y = 0$ the two asymmetries vanish (aperture positions are indicated by circles in Fig.~\ref{k_selection}~(c)). They also vanish for any position on the $k_y$ axis. An analogous reasoning holds for $y$-magnetization (``up''/``down'') by interchanging the Cartesian axes. Hence, depending on the aperture position one can selectively switch on/off the asymmetry signal of a particular magnetization direction. This momentum selection facilitates significantly the identification of domain structures.

\begin{figure}
	\includegraphics[width = \columnwidth]{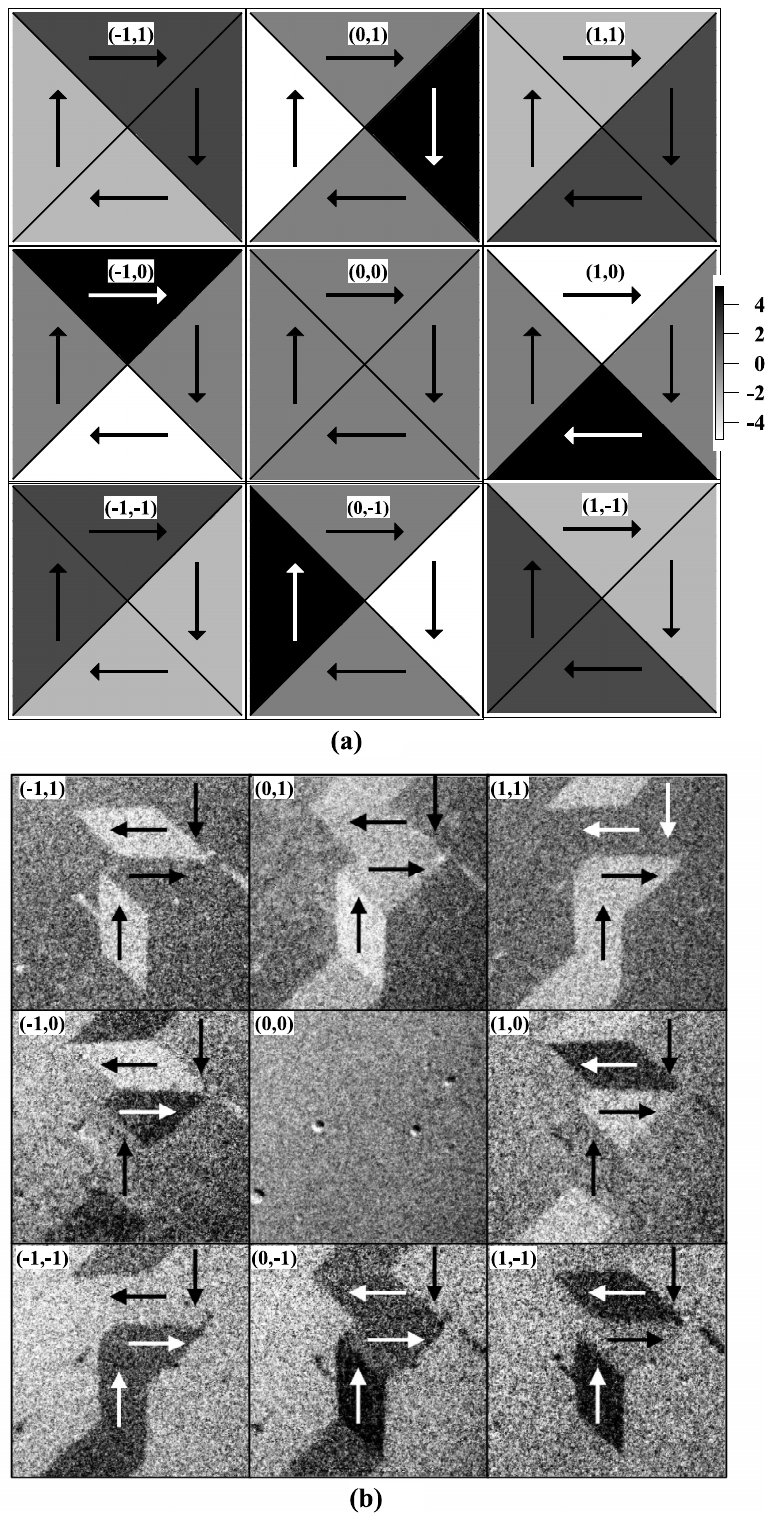}
	\caption{Panel (a) shows the simulated $A_{\mathrm{diff}}$ pattern of a 90$^{\circ}$ closure domain as a function of the momentum selection. The arrows indicate the magnetization direction. In panel (b) we display the experimental domain pattern. \label{domain_sim}}
\end{figure}

In order to illustrate this idea, we show how the asymmetries for a  $\unit[90]{^{\circ}}$-closure domain pattern change with position of the  contrast aperture (Fig.~\ref{domain_sim}(a)). The position of the nine images have a direct correspondence to the contrast aperture's position indicated in Fig.~\ref{k_selection}(c). 

The central image in Fig.~\ref{domain_sim}(a) refers to the momentum selection around $k_x = k_y = 0$, hence labeled $(0,0)$. The images on the left and right are from opposite selections with $k_y = 0$ and are termed $(-1,0)$ and $(1,0)$, respectively. The other images are to be understood analogously.

The $(0,0)$-image shows no contrast. Turning to the $(\pm 1,0)$-images one  can clearly separate between  the domains  ``right''  and ``left'', while the magnetization directions ``up''  and ``down'' can not be distinguished. Note that the signals for the ``right''  and ``left'' domain are reversed in the $(1,0)$- and the $(-1,0)$-images. The opposite domain sensitivity occurs for the $(0,\pm 1)$-images: this time the ``up''  and ``down'' domains can be discriminated in contrast to the ``right''  and ``left'' domains. 

Positioning the contrast aperture along a diagonal, e.g. image $(1,1)$ reveals that two pairs with orthogonal magnetization directions have the same contrast; for the $(1,1)$-image these are ``up'' -``right'' and ``down'' -``left''. This grouping remains the same for image $(-1,-1)$, only the contrast has changed its sign.

The concept outlined above has been applied in an experimental study of an Fe(001) surface (Fig.~\ref{domain_sim}(b)). The arrangement of the individual images reveals the central position of the contrast aperture, as in Fig.~\ref{domain_sim}(a). In the experiment, one does not know the orientation of the different magnetic domains within the image \textit{a priori}, but comparing the asymmetry pattern for different choices of the contrast aperture provides a unique answer.

 Starting with the central image $(0,0)$, no contrast is observed, as expected. Turning to the next image $(1,0)$ to the right, regions with different asymmetry values can be recognized. The two horizontal arrows indicate the magnetization directions of these two domains, labeled as ``right'' and ``left''. The two vertical arrows correspond to magnetization directions aligned along the $y$ axis, labeled as ``up'' and ``down''. Whether these are indeed oppositely magnetized cannot be determined at this point yet. Inspecting the image $(-1,0)$, located left of the center, shows that the contrast for the two horizontally magnetized domains (``right'' and ``left'') has reversed, while the areas marked by the vertical arrows maintain the same intensity.

We focus now on the image $(0,1)$ above the central image. Clearly the contrast between the  ``right''  and ``left' domains has vanished, while the regions marked with the vertical arrows are indeed oppositely magnetized domains  ``up''  and ``down''. Inspecting the images in the corners, e.g. $(1,1)$, we observe the same contrast for two pairs of domains. The magnetization direction within each of the pair is orthogonal.

\section{Sensitivity to perpendicular magnetization}
\label{sec:pma}
\begin{figure}
	\includegraphics[width = \columnwidth]{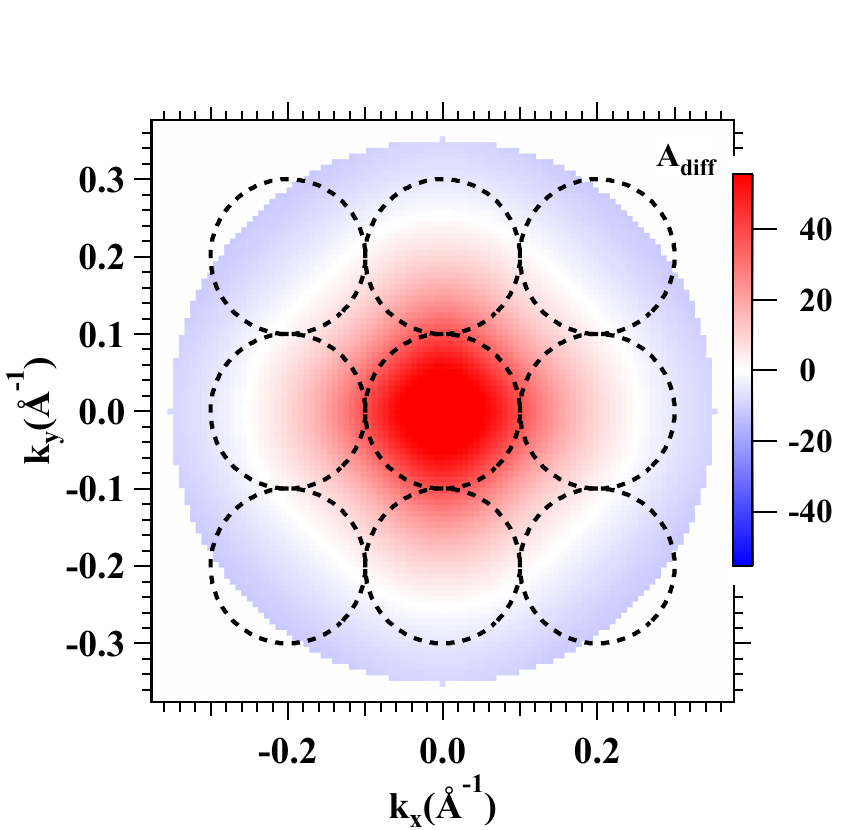}
	\caption{Difference asymmetry for perpendicular magnetization.  Data are calculated for normal incidence and $h \nu = \unit[5.20]{eV}$. Electrons are emitted from $E_F$. Circles  mark momentum ranges with a diameter of $\unit[0.2]{\AA^{-1}}$, given by the position of the contrast aperture analogous to Fig.~\ref{k_selection}(c).  \label{perp_mag}}
\end{figure}

In the discussion so far, we have neglected the possibility of a perpendicular orientation of the magnetization, which can occur in ultrathin films, such as Fe/Ag(001) films with a thickness below 5 ML \cite{2011_Ballentine,312_Qiu,541_Berger}. Here, we focus on the case of normally incident light.  Unlike to in-plane magnetization, emission from  $k_x = k_y = 0$ will result in a finite $A_{\mathrm{ex}}$. This is due to the symmetry properties of $A_{\mathrm{ex}}$ (cf.\ Eq.~\eqref{asym_trans2}). Since there are two orthogonal mirror planes, each component of the in-plane magnetization is either perpendicular to or lies within one of the mirror planes. As a result, there is a requirement to change or preserve the sign of $A_{\mathrm{ex}}$. This condition can only be satisfied if $A_{\mathrm{ex}}$ vanishes. However, for a perpendicular orientation, the magnetization lies within both the $xz$- and $yz$-planes, and thus $A_{\mathrm{ex}}$ generally adopts a finite value. To illustrate this, we have calculated $A_{\mathrm{diff}}$ for two oppositely magnetized perpendicular domains, as shown in Fig.~\ref{perp_mag} for $h \nu = \unit[5.20]{eV}$ and electron emission from  $E_F$. The central dashed aperture is centered at $k_x = k_y = 0$, where we observe a positive $A_{\mathrm{diff}}$ value in strong contrast to Fig. \ref{k_selection} (c).

In conclusion, with normally incident light, it is possible to select conditions that allow exclusive sensitivity to either one component of the in-plane magnetization or the perpendicular magnetization.

\section{Concluding Remarks}
\label{sec:conclusion}
We have investigated the photoemission process triggered by circularly polarized light on a ferromagnetic sample with $C_{\mathrm{4v}}$ symmetry. However, it suffices to consider the $C_{\mathrm{2v}}$ symmetry. The magnetic contrast arises from intensity differences formulated specifically for domain imaging. For a (001) surface, we examined how the experimental symmetry shapes the signal, leading to the idea of using a contrast aperture to target specific regions in the momentum space of the emitted electrons. By positioning the aperture properly, we achieve domain selectivity. Our experimental results are further supported by numerical simulations of the photoemission process from an Fe(001) surface, excited with $h\nu = \unit[5.20]{eV}$ (threshold photoemission). This surface has been experimentally studied using this photon energy, and the domain selectivity concept has been successfully confirmed.

\begin{acknowledgments}
This work is funded by the Deutsche Forschungsgemeinschaft (DFG, German Research Foundation) -- Project-ID 328545488 -- TRR~227, projects~A06 and~B04
\end{acknowledgments}

\appendix
\section{Asymmetry relations}
\label{sec:relations}
A complete measurement produces four intensities, $I_{\sigma,M}$, where the polarization state of the light $\sigma$ is alternated between positive ($\sigma_+$) and negative helicity ($\sigma_-$). Similarly, the magnetization direction is switched between two opposite orientations, labeled $M_{+}$ and $M_{-}$, respectively. Here, we introduce two alternative methods to compute a set of four expressions uniquely defined by the intensities $I_{\sigma,M}$.

The first method yields
\begin{subequations}
\begin{align} 
	\Delta_{\mathrm{pol}} & = I_{+,+}+  I_{+,-}- I_{-,+}-  I_{-,-},
	\label{d_pol} 
    \\
 \Delta_{\mathrm{mag}} & = I_{+,+}-  I_{+,-}+ I_{-,+}-  I_{-,-},  
	\label{d_mag}
    \\
 \Delta_{\mathrm{ex}} & = I_{+,+}-  I_{+,-}- I_{-,+}+  I_{-,-},  
	\label{d_ex}
    \\
 S & = I_{+,+}+I_{+,-}+ I_{-,+}+  I_{-,-}.  
	\label{S_tot}
\end{align}    
\end{subequations}
The first three equations comprise differences, while $S$ represents the sum of the intensities. Dividing the differences by $S$ yields the asymmetries as previously defined~\cite{2369_Henk}. The subscript ``pol'' refers to magnetization-averaged (polarization-related) dichroism, while ``ex'' identifies the exchange component of the dichroism. The dichroic signal obtained with unpolarized light corresponds to ``mag''~\cite{2369_Henk}.

In electron spectroscopy, the four intensities $I_{\sigma,M}$ can be measured by magnetizing the sample into oppositely magnetized single-domain states. However, this is not possible in domain imaging, as regions with different magnetization orientations are present, and only the helicity of the light can be reversed. Consequently, for each domain $M_{+}$ and $M_{-}$, labeled ``+'' and ``-'', a sum $S$ and a difference $\Delta$ of photoemission intensities can be defined as
\begin{subequations}
\begin{align} 
	\Delta_{+}= I_{+,+}- I_{-,+},
	\label{de_up}
    \\
 \Delta_{-}= I_{+,-}- I_{-,-}, 
	\label{de_down}
    \\
 S_{+}= I_{+,+}+ I_{-,+}, 
	\label{S_up}
    \\
 S_{-}= I_{+,-}+ I_{-,-}. 
	\label{S_down}
\end{align}    
\end{subequations}
With these definitions one can show that 
\begin{subequations}
\begin{align} 
	\Delta_{\mathrm{pol}}&= \Delta_{+}+\Delta_{-},
	\label{de_pol}
    \\
 \Delta_{\mathrm{mag}}&= S_{+}-S_{-}, 
	\label{de_mag}
    \\
 \Delta_{\mathrm{ex}}&= \Delta_{+}-\Delta_{-}, 
	\label{de_ex}
    \\
 S&= S_{+}+S_{-}. 
	\label{S_tot2}
\end{align}    
\end{subequations}

The above relations allow to rewrite the domain differences as
\begin{subequations}
\begin{align} 
	\Delta_{+} & =\frac{1}{2} (\Delta_{\mathrm{pol}}+\Delta_{\mathrm{ex}}),
	\label{de_up2}
    \\
    \Delta_{-} & = \frac{1}{2} (\Delta_{\mathrm{pol}}-\Delta_{\mathrm{ex}}).
	\label{de_down2}
\end{align}    
\end{subequations}
These equations are an important result, because it demonstrates that the dichroic signal from a domain has two contributions: it is not only determined by $\Delta_{\mathrm{ex}}$, but also by $\Delta_{\mathrm{pol}}$, which represents the magnetization-averaged contribution by definition. Notably, $\Delta_{\mathrm{mag}}$ does not contribute. 

In order to obtain the asymmetries $A_{+}$ and $A_{-}$, one has to divide $\Delta_{+}$ and $\Delta_{-}$ by either $S_{+}$ or $S_{-}$, respectively, while the asymmetries $A_{\mathrm{pol}}$, $A_{\mathrm{mag}}$, and $A_{\mathrm{ex}}$ are calculated by dividing by $S$. Since $\Delta_{\mathrm{mag}}$ is non-zero in general, $A_{\mathrm{mag}}$ does play a role. For small values of $A_{\mathrm{mag}}$, we arrive at
\begin{subequations}
\begin{align} 
	A_{+} & = (A_{\mathrm{pol}}+A_{\mathrm{ex}}) (1-A_{\mathrm{mag}}),
	\label{(A_+}
    \\
	A_{-} & = (A_{\mathrm{pol}}-A_{\mathrm{ex}})(1+A_{\mathrm{mag}}).
	\label{(A_-}
\end{align}
\end{subequations}
This implies that the asymmetry signal from a ``+'' or ``$-$'' domain is primarily determined by the asymmetries $A_{\mathrm{pol}}$ and $A_{\mathrm{ex}}$, while $A_{\mathrm{mag}}$ acts as a correction. Furthermore, if $A_{+}$ and $A_{-}$ have the same sign, $A_{\mathrm{pol}}$ is larger in magnitude than $A_{\mathrm{ex}}$. Particularly relevant for domain imaging is the difference between $A_{+}$ and $A_{-}$,
\begin{align} 
	A_{\mathrm{diff}} & =
 2 \, \left( A_{\mathrm{ex}}-A_{\mathrm{pol}}A_{\mathrm{mag}} \right) \approx 2 \, A_{\mathrm{ex}}.
	\label{(A_diff}
\end{align}

%

\end{document}